# The Electronic States of a Double Carbon Vacancy Defect in Pyrene: A Model Study for Graphene

Francisco B. C. Machado,*,[1] Adélia J. A. Aquino[2,3] and Hans Lischka*,[2,3]

[1] Departamento de Química, Instituto Tecnológico de Aeronáutica, São José dos Campos, 12228-900, São Paulo, Brazil

[2] Department of Chemistry and Biochemistry, Texas Tech University Lubbock, TX 79409-1061 (USA)

[3] Institute for Theoretical Chemistry, University of Vienna, A-1090 Vienna, Austria

E-mail: fmachado@ita.br, hans.lischka@univie.ac.at

**Abstract**

The electronic states occurring in a double vacancy defect for graphene nanoribbons have been calculated in detail based on a pyrene model. Extended *ab initio* calculations using the MR configuration interaction (MRCI) method have been performed to describe in a balanced way the manifold of electronic states derived from the dangling bonds created by initial removal of two neighboring carbon atoms from the graphene network. In total, this study took into account the characterization of 16 electronic states (eight singlets and eight triplets) considering unrelaxed and relaxed defect structures. The ground state was found to be of $^1A_g$ character with around 50% closed shell character. The geometry optimization process leads to the formation of two five-membered rings in a pentagon–octagon–pentagon (5–8–5) structure. The closed shell character increases thereby to ~70%; the analysis of unpaired density shows only small contributions confirming the chemical stability of that entity. For the unrelaxed structure the first five excited states ($^3B_{3g}$, $^3B_{2u}$, $^3B_{1u}$, $^3A_u$ and $^1A_u$) are separated from the ground state by less than 2.5 eV. For comparison, unrestricted density functional theory (DFT) calculations using several types of functionals have been performed within different symmetry subspaces defined by the open shell orbitals. Comparison with the MRCI results gave good agreement in terms of finding the $^1A_g$ state as ground state and in assigning the lowest excited states. Linear interpolation curves between the unrelaxed and relaxed defect structures also showed good agreement between the two classes of methods opening up the possibilities of using extended nanoflakes for multistate investigations at DFT level.

## Introduction

Since the initial discovery of graphene[1, 2] it quickly became clear that this was one of the most promising materials in the quest for future nanoscale technologies. Graphene consists of a single atomic layer of graphite and possesses exceptional electronic, thermal and mechanical properties. Its promising applications in electronics, optoelectronics and photonics have been described in many articles.[3-6] Graphene is a semimetal and the absence of a band gap limits graphene to be used as electronic device. The introduction of defects into the regular honeycomb network of graphene constitutes an important technique to modify the graphene properties. To achieve this goal in a controlled way the characterization of the electronic structure of the defect states, their structures and energetics is of great significance.

Vacancy defects represent an important class of structural features where carbon atoms are missing in the hexagonal structure of graphene and dangling bonds introducing high chemical reactivity occur. These defects arise in graphene or in graphitic nanostructures during defective growth and can also be created artificially by means of ion irradiation.[7-14] Their structural details can be directly observed by means of several experimental techniques such as transmission electron microscopy (TEM)[15-19] and scanning tunneling microscopy (STM).[20, 21]

Because of the occurrence of dangling bonds the defect structure will be associated with a high polyradical character with a multitude of closely spaced locally excited electronic states possessing different spin multiplicities which make their theoretical description very challenging. For such situations multireference theory[22] provides flexible and efficient tools to compute and analyze these electronic states by combining sets of quasi-degenerate orbitals at equal footing allowing the construction of appropriate wavefunctions of well-defined symmetry and spin properties.

Recently, we have shown[23] by means of multireference configuration interaction (MRCI) calculations that a single vacancy (SV) defect induces a complex set of several closely spaced electronic states leading to geometry relaxation effects with carbon-carbon bond formation (Scheme 1b and c). This bond formation had been predicted based on the analysis of Jahn-Teller symmetry breaking,[24] which was confirmed later on by density functional calculations.[25-29] In a first approach, pyrene (Scheme 1a) has been used since it contains the basic structural body describing the vacancy defect. The MRCI calculations performed showed that the complexity of the electronic states was in fact significantly more pronounced than anticipated in the previous work and that avoided crossings between different states occurred which transformed an originally antibonding character into a bonding one; as a consequence this behavior led to bond reconstruction formation not only for one state, but actually for several ones including singlet and triplet states.

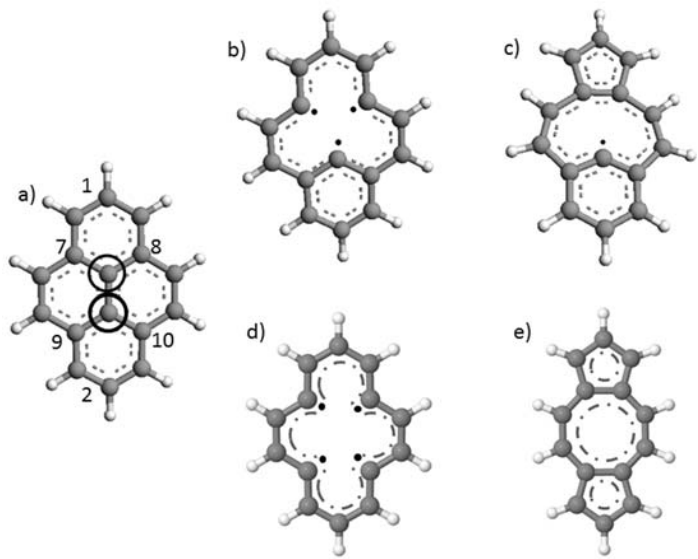

**Scheme** 1. a) pyrene with numbering of selected carbon atoms, the circles indicate the carbon atoms to be removed, b) pyrene-1C-unrel c) pyrene-1C-relaxed, d) pyrene-2C-unrel, e) pyrene-2C-relaxed.

Based on the afore-mentioned pyrene model, in the present study the properties of a double vacancy (DV) are investigated. Scheme 1a shows the original unperturbed structure. In Scheme 1d two carbon atoms have been replaced instantaneously creating four dangling bonds in the σ orbital system. The original unrelaxed structure is still retained. In Scheme 1e geometry relaxation has occurred. The basic features of this defect have been explained by Coulson et al.[24, 30] many years ago on the basis of a tight binding model taking into account the π system and the four dangling bonds of the σ orbitals. More detailed investigations on the electronic structure of graphene with double vacancies have been performed later on.[25-29, 31-39] Most of these calculations found a pentagon–octagon–pentagon (5–8–5) structure similar to the prototype in Scheme 1e and as suggested from an analysis of Raman scattering spectra of irradiated graphite.[7] The newly formed σ bond length of each pentagon calculated using density functional theory (DFT) within the local density approximation (LDA)[36] is equal to 1.77 Å. In spite of the numerous theoretical studies on the graphene double vacancies, the characterization of the electronic state manifold is still lacking attention.

The main objective of the present work is to investigate in detail the electronic states of the double vacancy defect originating from the dangling bonds of the σ system together with the coupling to the π orbitals. Introduction of a double vacancy enhances the complexity of the computational problem considerably as compared to the single vacancy because four dangling bonds are created by removal of the two carbon atoms (compare Scheme 1b with Scheme 1d). Similar to our previous investigation on the single vacancy, the MRCI approach will be used. As has already been observed for the SV defect,[23] several electronic states with different spin multiplicities are to be expected which will in part be closely spaced depending on coupling of the different open shell orbitals. The MRCI calculations are computationally expensive. On the other hand the reliability of cheaper methods in terms of computer time, especially of density functional theory (DFT) is not clear in view of the many electronic states to be investigated. Therefore, the second goal of this work is to assess the applicability of DFT for the present DV case taking the MRCI results as benchmark.

## Methods

Complete active space (CAS) self-consistent field (CASSCF)[40, 41] and MRCI[22] calculations have been performed on the double vacancy structures shown in Scheme 1d and e. A CAS (8,8) with eight electrons and eight orbitals was chosen for the CASSCF calculations using one orbital for each irreducible representation. In $D_{2h}$ symmetry this set of molecular orbitals (MOs) was constructed from the $12a_g$, $11b_{1u}$, $2b_{1g}$, $2b_{2g}$, $3b_{3u}$ $10b_{2u}$, $2a_u$ and $9b_{3g}$ orbitals, which are, respectively, the four highest occupied orbitals and the four lowest virtual orbitals of the $^1A_g$ state at density functional theory (DFT) level using the B3-LYP[42-44] functional and the 6-31G** basis.[45] The shape and functionality of these orbitals will be discussed in the Results section. These MOs were used in the MRCI calculations with a CAS(8,8) reference space identical to that one employed in the CASSCF calculations. Single and double excitations were constructed from the occupied orbitals into the entire virtual orbital space applying the interacting space restriction;[46] only the 1s carbon orbitals were kept frozen. Size-extensivity contributions are included by means of the Davidson correction[22, 47] which is denoted by the label +Q (MRCI+Q).

The 6-31G and 6-31G* basis sets were used throughout the calculations.[45] The original pyrene structure was obtained from DFT/B3-LYP optimizations using the 6-31G** basis. The two innermost carbon atoms (Scheme 1a) were then removed. The resulting structure is denoted pyrene-2C-unrel. CASSCF/6-31G* geometry optimizations were performed for each electronic state separately to obtain the pyrene-2C-relaxed structures. Linear interpolation curves between the unrelaxed and relaxed structure were computed at the MRCI+Q (8,8) and DFT/B3-LYP levels using the 6-31G* basis set. The pyrene molecule was arranged in the yz plane with the long axis oriented along the z axis. Unrestricted (U)DFT/B3-LYP single point calculations and geometry optimizations were performed using the 6-31G* basis set. Singlet and triplet multiplicities were considered. The state symmetry in the UDFT calculations was determined from the direct product of the irreducible representations of the respective open shell orbitals. The occupation schemes for the DFT calculations are presented in Table 6S of the Suplementary Material. All optimized structures preserved the $D_{2h}$ point group symmetry.

The effective unpaired electron densities and total number of effectively unpaired electrons ($N_U$) were computed[48-50] in order to characterize the polyradical character of the different states. To avoid overemphasizing the contribution of the natural orbitals (NOs) that are nearly doubly occupied or nearly unoccupied, we chose to use the non-linear model suggested in Ref.[44] where $N_U$ is given by

$$N_U = \sum_{i=1}^{M} n_i^2 (2-n_i)^2 \qquad (1)$$

in which $n_i$ is the occupation of the $i$-th NO, and $M$ is the number of NOs.

The MR calculations were performed with the COLUMBUS program system,[51-53] using its parallel version.[54,55] CASSCF geometry optimizations were performed with the program system DL-FIND[56] interfaced to COLUMBUS. For the DFT calculations the TURBOMOLE program[57] was used.

## Results and Discussion

### Unrelaxed Structure

The vertical excitations for the first six states of the pyrene-2C-unrel structure calculated using the CASSCF, MR-CISD and MR-CISD+Q and B3-LYP methods are collected in Table **1** together with the characterization of the main configuration. In total sixteen electronic states have been calculated, considering singlet and triplet spin multiplicity for each of the irreducible representations in $D_{2h}$ symmetry. The results for all calculated sixteen states obtained with the 6-31G* and 631G basis set are collected in Tables 1S and 2S of the Supplementary Information. The Cartesian coordinates of the unrelaxed structure is given in the Supplementary Information as well. The natural orbital occupations for the most important orbitals are displayed in Table 3S for the six low-lying states. The ground state has symmetry $^1A_g$ and the first five excited states which lie above the ground state by less than 3.0 eV calculated at MR level are $^3B_{3g}$, $^3B_{2u}$, $^3B_{1u}$, $^3A_u$ and $^1A_u$. It is also important to note that there is only a weak dependence of the excitation energies on the computational method (CASSCF, MRCI or MRCI+Q) and on the basis sets used (Table 1 and Table 1S). The weight of the dominating configuration is less than 60% for all states computed, which is an indication of the multiconfigurational character of all states. This fact is also exemplified e.g. for the $^1A_g$ state where afore-mentioned NO occupations for the occupied σ bonding orbitals ($12a_g$ and $11b_{1u}$) and the unoccupied σ* antibonding orbitals ($10b_{2u}$ and $9b_{3g}$) values of around 1.7 and 0.3, respectively, are found (Table S3). These values deviate substantially from the standard closed shell occupations. The DFT/B3-LYP calculations also find the $^1A_g$ as ground state. The first excited state is $^3A_u$, ~1.0 eV above the $^1A_g$ ground state. The order of the excited states computed at the DFT level differs from that of the MR calculations (see also Table 1S) since these states present a strong multiconfigurational character. Nevertheless, the set of lowest singlet and triplet states is the same in the MR and DFT calculations, in spite of the just-mentioned differences in the detailed energetic ordering. The calculations were also carried out using PBE[58] and PBE0[59] functionals, which present the same electronic excitation ordering and similar excitation energies, differing no more than 0.2 eV. These results are presented in Table 7S.

**Table** 1. Excitation energies (eV) for the unrelaxed pyrene-2C structure using a CAS (8,8) reference space and B3-LYP, respectively, together with the 6-31G* basis set.

| State | CASSCF | MRCI | MRCI+Q | B3-LYP | Config.[a,b] |
|---|---|---|---|---|---|
| $^1A_g$ | 0.000[c] | 0.000[c] | 0.000[c] | 0.000[c] | $12a_g^2 3b_{3u}^0 10b_{2u}^0 2b_{1g}^2 11b_{1u}^2 2b_{2g}^2 9b_{3g}^0 2a_u^0$ (54%) |
| $^3B_{3g}$ | 1.607 | 1.635 | 1.646 | 1.616 | $12a_g^2 3b_{3u}^0 10b_{2u}^1 2b_{1g}^2 11b_{1u}^1 2b_{2g}^2 9b_{3g}^0 2a_u^0$ (41%) |
| $^3B_{2u}$ | 1.741 | 1.799 | 1.827 | 2.378 | $12a_g^2 3b_{3u}^0 10b_{2u}^0 2b_{1g}^2 11b_{1u}^1 2b_{2g}^2 9b_{3g}^1 2a_u^0$ (31%) |
| $^3B_{1u}$ | 1.774 | 1.752 | 1.712 | 1.405 | $12a_g^2 3b_{3u}^1 10b_{2u}^0 2b_{1g}^2 11b_{1u}^2 2b_{2g}^1 9b_{3g}^0 2a_u^0$ (57%) |
| $^3A_u$ | 2.905 | 2.524 | 2.330 | 1.057 | $12a_g^2 3b_{3u}^0 10b_{2u}^1 2b_{1g}^2 11b_{1u}^2 2b_{2g}^1 9b_{3g}^0 2a_u^0$ (58%) |
| $^1A_u$ | 2.980 | 2.618 | 2.433 | 1.181 | $12a_g^2 3b_{3u}^0 10b_{2u}^1 2b_{1g}^2 11b_{1u}^2 2b_{2g}^1 9b_{3g}^0 2a_u^0$ (55%) |

[a] Closed shell part: $11a_g^2 2b_{3u}^2 9b_{2u}^2 1b_{1g}^2 10b_{1u}^2 1b_{2g}^2 8b_{3g}^2 1a_u^2$.

[b] MRCI configuration percentage in parentheses.

[c] Total energies (hartree): $^1A_g$ CASSCF/6-31G* = -535.6751476 (-14576.468 eV); $^1A_g$ MRCI/6-31G* = -536.9015422 (-14609.840 eV); $^1A_g$ MRCI + Q/6-31G* = -537.1933974 (-14617.782 eV); B3-LYP/6-31G* = -538.7885053 (-14661.18722 eV).

The active CASSCF molecular orbitals computed for the $^1A_g$ state are displayed in Figure 1. With respect to the $C_7$-$C_8$ and $C_9$-$C_{10}$ bonds (Scheme 1d) which are formed on geometry relaxation, the $12a_g$ and $11b_{1u}$ orbitals are σ bonding, the $10b_{2u}$ and $9b_{3g}$ are σ* antibonding, the π orbitals $2b_{2g}$ and $3b_{3u}$ are bonding and $2b_{1g}$ and $2a_u$ are π* antibonding. The B3-LYP orbitals are quite similar to the CASSCF orbitals (see Figure 1S in the Supporting Information). This analysis shows (see Table 1) that the first two excited states, $^3B_{3g}$, and $^3B_{2u}$, arise from the σ-σ* excitations, the $^3A_u$ and $^1A_u$ states from π-σ* excitations and the $^3B_{1u}$ state results from a π-π excitation. Thus, the four former excited states acquire CC antibonding character on electronic excitation from the $^1A_g$ ground state.

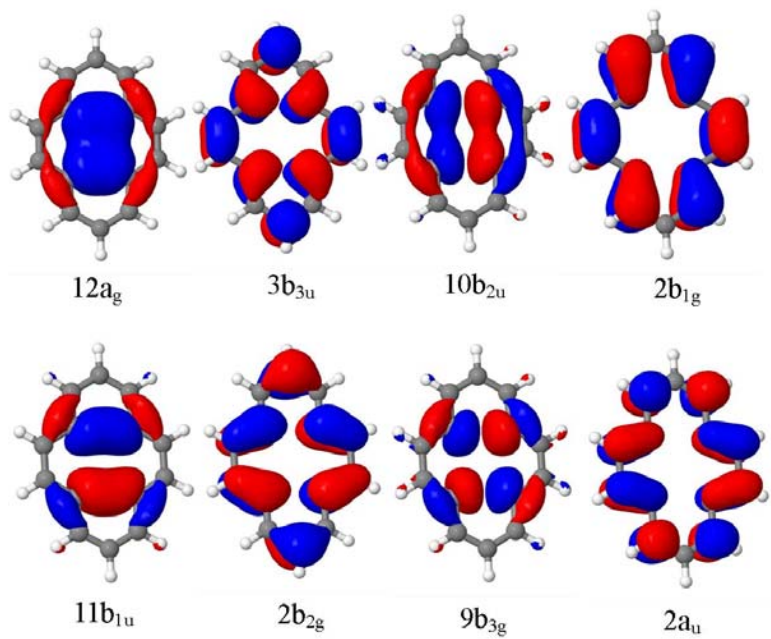

**Figure 1**. Active molecular orbitals for the $^1A_g$ state computed at the CASSCF(8,8)/6-31G* level for the unrelaxed structure.

## Geometry relaxation

Geometry optimizations have been performed at the CASSCF(8,8) and B3-LYP levels for the low-lying states ($^1A_g$, $^3B_{2u}$, $^3B_{1u}$, $^1B_{2u}$) as identified in the linear interpolation calculations discussed below. At the optimized CASSCF geometries single point MRCI calculations were carried out. Table 2 collects the relative stabilities, the optimized $C_7$-$C_8$ distance and the main electronic configuration. The results for the 6-31G basis set (Table 4S), the natural orbital occupations (Table 5S) and the Cartesian geometries can be found in the Supporting Information. For all states calculated a strong reduction of the $C_7$-$C_8$ ($C_9$-$C_{10}$) distances from originally 2.47 Å in the unrelaxed pyrene-2C-unrel structure to distances between 1.45 Å to 1.54 Å is observed. This range of distances covers standard CC single bonds lengths and somewhat shortened ones. Similar results are obtained at CASSCF and DFT/B3LYP level. They indicate the formation of the pentagon–octagon–pentagon (5–8–5) structure shown in Scheme 1d. For comparison, local density approximation (LDA)[36] calculations using periodic boundary conditions find a value of 1.77 Å for the $C_7$-$C_8$ bond length. The $^1A_g$ B3-LYP and CASSCF results of 1.52 and 1.54 Å, respectively, are smaller which is due to the larger structural flexibility of our pyrene model. For the relaxed structures the same energetic ordering of the four low-lying states shown in Table 2 is obtained at CASSCF, MRCI and B3-LYP levels: the ground state has $^1A_g$ symmetry followed by the $^3B_{2u}$, $^1B_{2u}$ and $^3B_{1u}$ states. At MRCI+Q level the lowest excited state ($^3B_{2u}$) is 1 eV above the ground state and the next ones are following in distances of several tenths of an eV. In comparison, the B3-LYP excitation energies are somewhat smaller. The two lowest excited states are only 0.4 to 0.5 eV above the ground state. Note that the weight of the dominating configuration is ~70% or more at MRCI level for all states computed – a number which is better suited for the characterization of the respective state than the <60% reported for the unrelaxed structure discussed above. Again, we find a relatively small effect of increasing the basis set on energy differences and geometries (cf. Table 2 and Table 4S). The four low-lying states were also optimized using PBE and PBE0 functionals, presenting results similar to B3-LYP (see Table 8S).

Table 2: Excitation energies ΔE (eV) and optimized $C_7$-$C_8$ distance (Å)[a] for the relaxed pyrene-2C structure using a CAS (8,8) reference space and B3LYP with the 6-31G* basis set.

| State | CASSCF | | MRCI[b] | MRCI+Q[b] | B3-LYP | | Config.[c,d] |
|---|---|---|---|---|---|---|---|
| | ΔE | $C_7$-$C_8$ | ΔE | ΔE | $C_7$-$C_8$ | ΔE | |
| $^1A_g$ | 0.000[e] | 1.535 | 0.000[e] | 0.000[e] | 1.515 | 0.000[e] | $12a_g^2 3b_{3u}^0 10b_{2u}^0 2b_{1g}^2 11b_{1u}^2$ $2b_{2g}^2 9b_{3g}^0 2a_u^0$ (69%) |
| $^3B_{2u}$ | 1.182 | 1.472 | 1.102 | 1.053 | 1.454 | 0.441 | $12a_g^2 3b_{3u}^1 10b_{2u}^0 2b_{1g}^1 11b_{1u}^2$ $2b_{2g}^2 9b_{3g}^0 2a_u^0$ (75%) |
| $^1B_{2u}$ | 1.907 | 1.467 | 1.540 | 1.323 | 1.453 | 0.479 | $12a_g^2 3b_{3u}^1 10b_{2u}^0 2b_{1g}^1 11b_{1u}^2$ $2b_{2g}^2 9b_{3g}^0 2a_u^0$ (73%) |
| $^3B_{1u}$ | 1.535 | 1.520 | 1.626 | 1.630 | 1.502 | 1.430 | $12a_g^2 3b_{3u}^1 10b_{2u}^0 2b_{1g}^2 11b_{1u}^2$ $2b_{2g}^1 9b_{3g}^0 2a_u^0$ (75%) |

[a] $C_7$-$C_8$ and $C_9$-$C_{10}$ distances are symmetry equivalent.
[b] Single point calculation at CASCSF (8,8) optimized geometries.
[c] Closed shell part: $11a_g^2 2b_{3u}^2 9b_{2u}^2 1b_{1g}^2 10b_{1u}^2 1b_{2g}^2 8b_{3g}^2 1a_u^2$.
[d] MRCI configuration percentage in parentheses.
[e] Total energies (hartree): $^1A_g$ CASSCF/6-31G* = -535.9117919 (-14582.908 eV); $^1A_g$ MRCI/6-31G* = -536.8162795 (-14607.520 eV); $^1A_g$ MRCI + Q/6-31G* = -537.4457100 (-14624.64788 eV); $^1A_g$ B3-LYP/6-31G* = -539.05378069 (-14668.406 eV).

The active CASSCF orbitals computed for the relaxed structure of the $^1A_g$ state are displayed in Figure 2. The B3-LYP orbitals are similar to the CASSCF orbitals as one can see from Figure 1S. As compared to the unrelaxed structure, the formation of the two new bonds ($C_7$-$C_8$ and $C_9$-$C_{10}$) concentrates the bonding character of the σ bonding orbitals ($12a_g$ and $11b_{1u}$) and the π bonding orbitals ($2b_{2g}$ and $3b_{3u}$) significantly. Note also a reordering of the excited states as compared to the unrelaxed defect structure (Table 1). The $^3B_{2u}$ and $^1B_{2u}$ states are now the first and second excited states. In comparison to the orbital occupations found for the $^3B_{2u}$ state in the unrelaxed structure (Table 1) a significant reorganization of orbital occupations has occurred (Table 2). Instead of the original σ–σ* excitation ($11b_{1u} \rightarrow 9b_{3g}$), this state arises now from a π–π excitation from the antibonding $2b_{1g}$ into the bonding $3b_{3u}$ orbital (Figure 2). The $^1B_{2u}$ state possesses the same π–π excitation into the bonding $3b_{3u}$ orbital. The $^3B_{1u}$ state arises from an excitation from a bonding orbital ($2b_{2g}$) to another bonding orbital ($3b_{3u}$). The singlet and triplet $B_{2u}$ states present the shortest $C_7$-$C_8$ and $C_9$-$C_{10}$ bond distances, followed by the $^3B_{1u}$, while the ground state ($^1A_g$) has the largest distance. Note that at the ground state ($^1A_g$) the bonding orbital ($3b_{3u}$) is not occupied. The $^3B_{3g}$, $^3A_u$ and $^1A_u$ states listed in Table 1 have CC antibonding character and, thus, are not stabilized (see the section on Linear Interpolation Curves below). The optimized bond lengths of the formed σ bonds as displayed in Table 2 are all in the range of 1.50 ± 0.03 Å. For comparison, DFT/LDA calculations[36] on a double vacancy in a repeating slab model give a respective CC distance of 1.77 Å which demonstrates the larger rigidity of the graphene network as compared to the present pyrene model.

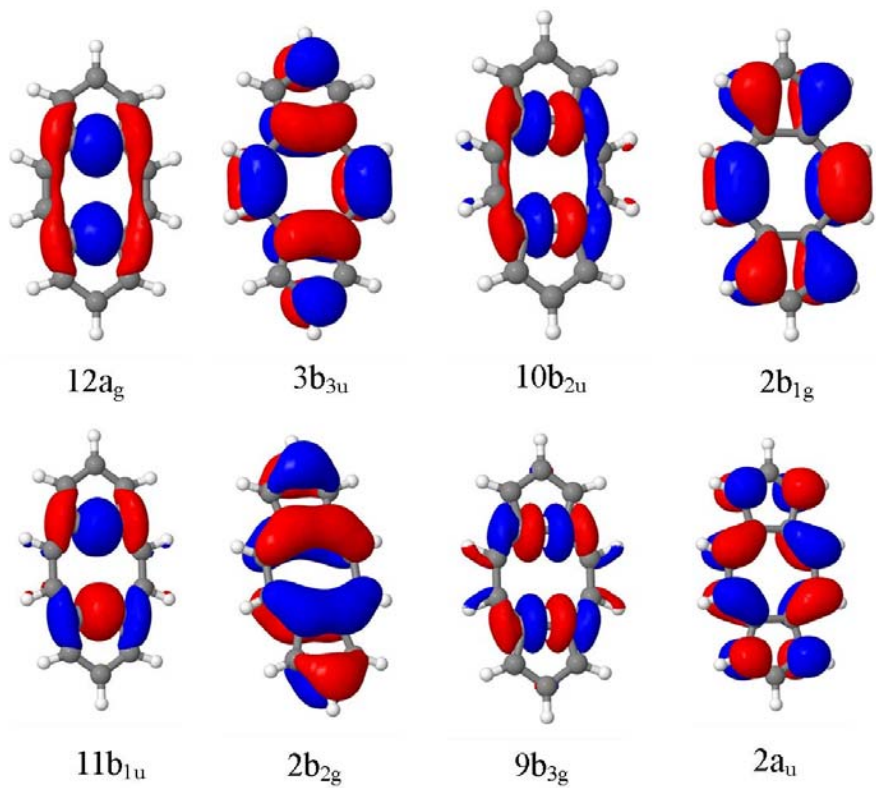

**Figure 2.** Active molecular orbitals for the $^1A_g$ state computed at the CASSCF(8,8)/6-31G* level for the relaxed structure.

## Linear Interpolation Curves

To analyze the evolution of the electronic character of the different states, linear interpolation curves connecting the structures with $C_7$-$C_8$ and $C_9$-$C_{10}$ from 2.744 Å (unrelaxed) to 1.467 Å (relaxed, $^1B_{2u}$ state) were computed at MRCI+Q level for each of those geometries in a total of sixteen electronic states. In Figure 3 and Figure 4 the linear interpolation curves for the first four low-lying electronic states are presented using MRCI+Q and DFT/B3-LYP, respectively. In Figure 2S the MRCI+Q linear interpolation curves for the total of sixteen states are given. All states shown in Figure 3 and Figure 4 are stabilized by decreasing the $C_7$-$C_8$ and $C_9$-$C_{10}$ bond distance. It is noted that comparison of the MRCI+Q and DFT/B3-LYP curves displayed in the two figures show very good agreement. The overall stabilization energy with respect to the unrelaxed defect structure is substantial and amounts to more than 5.0 eV at MRCI+Q level. As already mentioned above, the $^3B_{2u}$ state changes occupation which happens at a $C_7$-$C_8$ distance of ~2.3 Å because of an avoided crossing with a higher $B_{2u}$ state. Inspection of the linear interpolation curves for the higher excited states (Figure 2S) shows that many other states also change occupation and stabilize at short a $C_7$-$C_8$ bond distance. The $^3B_{3g}$ state, which is the first excited state at the unrelaxed structure, remains repulsive until ~2.05 Å where it is stabilized by an avoided crossing, but still stays about 3.8 eV above the ground state.

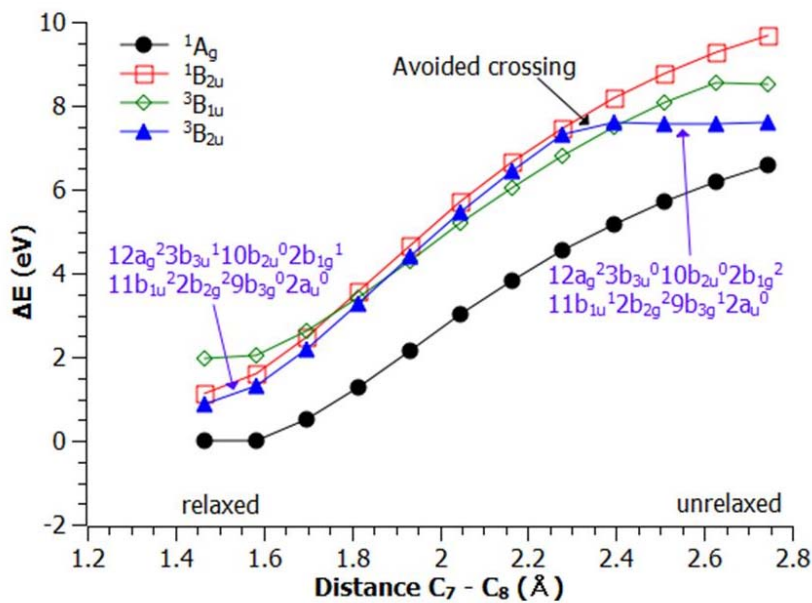

**Figure 3.** Linear interpolation curve for the first four low-lying states of pyrene-2C computed at MRCI + Q (8,8)/6-31G* level. Paths between the structures with $C_7$-$C_8$ and $C_9$-$C_{10}$ = 2.744 Å (unrelaxed) to 1.467 Å (relaxed $^1B_{2u}$ state). Energies are relative to $E(^1A_g)$ = -537.4386964 hartree (-14624.457 eV). The main electronic configuration of the $^3B_{2u}$ state for the unrelaxed and relaxed structure is given in blue.

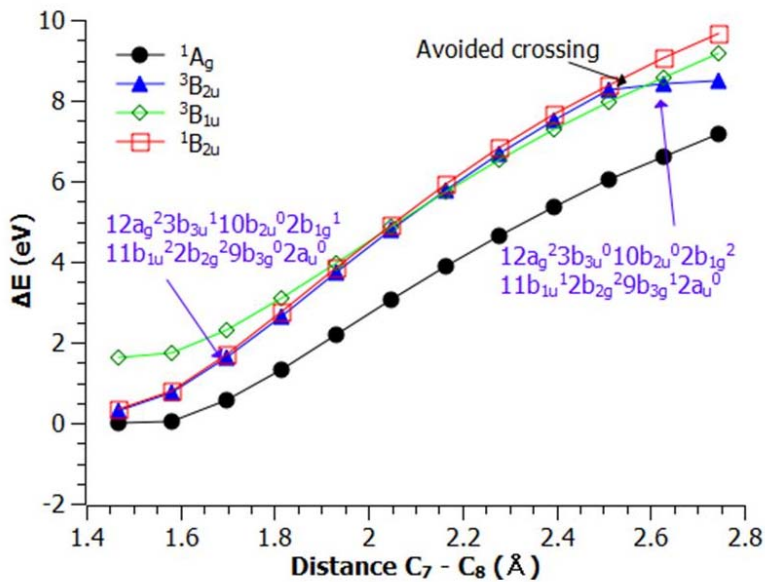

**Figure 4**. Linear Interpolation for the first four low-lying states of pyrene-2C computed at DFT/B3-LYP/6-31G* level. Paths between the structures with $C_7$-$C_8$ and $C_9$-$C_{10}$ = 2.744 Å (unrelaxed) to 1.467 Å (relaxed $^1B_{2u}$ state). Energies are relative to E($^1A_g$) = -539.0457677 hartree (-14668.188 eV). In blue, the main electronic configuration of the $^3B_{2u}$ state along the path.

Unpaired densities

The unpaired densities displayed in Figure 5 for the unrelaxed and relaxed defect structures summarize the electronic structures of the different electronic states. For the $^1A_g$ state Figure 5a shows that for the unrelaxed structure the radical character is almost exclusively located in the region of the $C_7$-$C_8$ and $C_9$-$C_{10}$ bonds. For the relaxed structure (Figure 5d), where the new bonds are formed, the total number of unpaired density $N_U$ is practically zero ($N_U$ = 0.20$e$) which illustrates nicely the closed shell character and the chemical stability of this state. For the unrelaxed structure of the $^3B_{2u}$ state the multiradical character is concentrated in the region of the $C_7$-$C_8$ and $C_9$-$C_{10}$ bonds (Figure 5b). For the relaxed structure (Figure 5e), the same electronic state possesses a completely different appearance as is to be expected from the above discussion of the avoided crossing and the change in the character of the electronic wavefunction on the CC bond formation. There is some unpaired π density located in the two five-membered rings and in the octagon of the pentagon–octagon–pentagon (5–8–5) structure. The same shape of unpaired density is found also for the almost degenerate relaxed $^1B_{2u}$ state (see Figure S3b in Supplementary Information). In both the unrelaxed and relaxed structure the $^3B_{1u}$ state (Figure 5 c,f) possesses unpaired π densities located in the region of the $C_7$-$C_8$ and $C_9$-$C_{10}$ bonds and also shows unpaired π density located at $C_1$ and $C_2$. The total number of unpaired density $N_U$ reduces significantly for all cases shown in Figure 5 when going from the unrelaxed to the relaxed structure. For the latter structure, the $^1A_g$ state does not possess any significant radical character whereas values of ~2*e* are observed for the other states indicating significant biradical character. This discussion shows that for the relaxed DV defect structure the electronic ground state is mainly of closed shell character and relatively stable. However, starting with about 1 eV above this state (Table 2) highly reactive states are located which could be accessible e.g. in chemisorption processes.

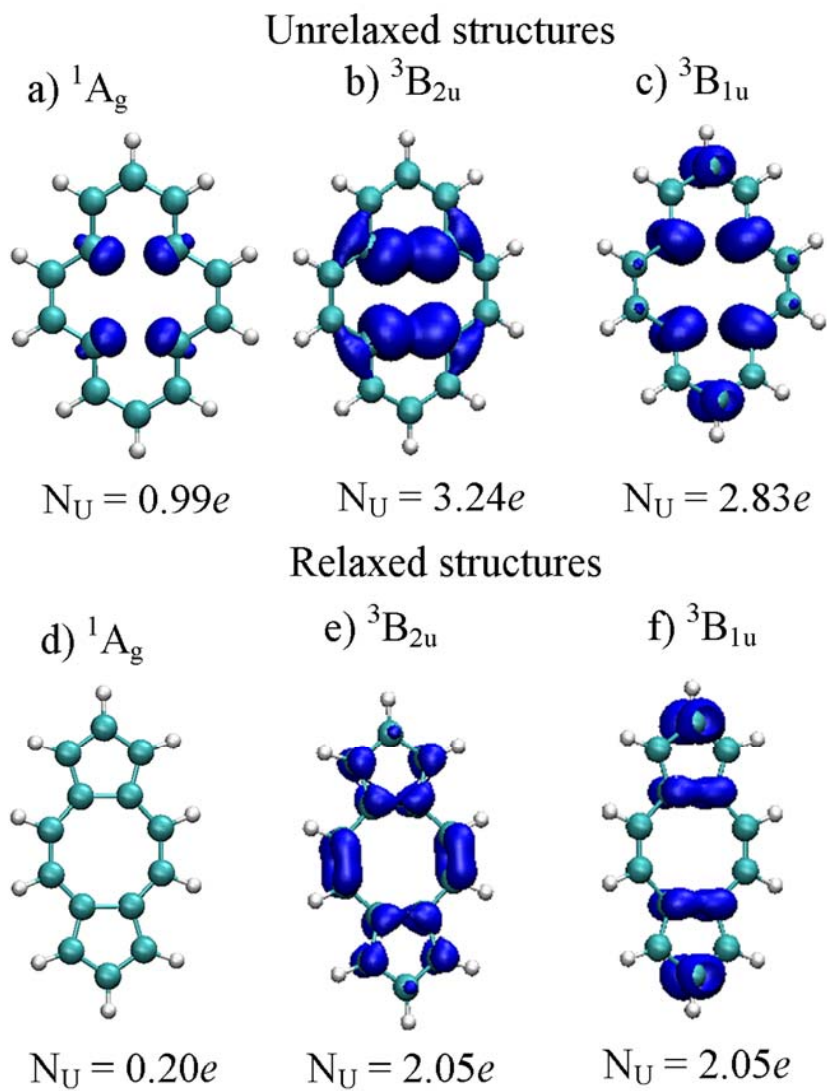

**Figure 5**. Unpaired electron density plots for the $^1A_g$, $^3B_{2u}$ and $^3B_{1u}$ states using the MR-CISD/6-31G* approach. a), b) and c) unrelaxed geometry, d), e) and f) relaxed geometry. Isodensity value is 0.007 $e$/bohr$^3$.

## Conclusions

The diversity of the electronic states formed by a double vacancy defect in a graphene nanosheet has been studied using pyrene as a model structure using *ab initio* CASSCF and MRCI+Q as well as DFT/B3-LYP calculations. In total sixteen electronic states, eight singlets and eight triplets, were computed. The ground state has symmetry $^1A_g$. The unrelaxed structure originating from pristine pyrene and geometry relaxed structures were investigated. For the unrelaxed structure the first five excited states ($^3B_{3g}$, $^3B_{2u}$, $^3B_{1u}$, $^3A_u$ and $^1A_u$) are separated from the ground state by less than 2.5 eV. The DFT calculations also find the $^1A_g$ state as ground state. However, a slightly different order of the excited states is found, presumably since these states possess a strong multiconfigurational character for which this method has not been adapted for. For the relaxed structures the first low-lying states ($^1A_g$, $^3B_{2u}$, $^3B_{1u}$, $^1B_{2u}$) have the same energetic ordering in MRCI+Q and DFT/B3-LYP. The geometry optimization process leads to a strong reduction of the $C_7$–$C_8$ and $C_9$–$C_{10}$ distances forming the two five-membered rings in a pentagon–octagon–pentagon (5–8–5) structure. In this bond formation process the triplet $B_{2u}$ state changes its configuration due an avoided crossing and its unpaired π density becomes located at the short CC bond on the two five-membered rings and on the octagon. At the relaxed structure the $^1A_g$ ground state is dominated by a closed shell configuration with almost negligible unpaired electron density. This stability distinguishes the double vacancy from the single vacancy defect where we have found previously[37] that even for the relaxed structure a substantial unpaired density existed primarily due to the occurrence of a dangling bond.

Currently, we have obtained a detailed picture of the manifold of electronic states occurring due to the removal of two carbon atom from a graphene sheet based on the local environment of the defect. The multireference methods used allow a general treatment of the complicated electronic coupling processes occurring between the different unpaired electrons in the defect and the coupling to the π system. Building on the experience gained in the present calculations, especially with respect of the good agreement between the MRCI and DFT results, the investigation of significantly larger graphene sheet models will be accessible providing better insight into the embedding effects and their consequences on the electronic and geometrical structure of the double vacancy defect.

Electronic Supplementary Information (ESI) available: Full sets of excitation energies calculated, molecular orbitals, natural orbital occupations, unpaired electron density plots, DFT occupation schemes, and Cartesian coordinates.

# ACKNOWLEDGMENTS

This material is based upon work supported by the National Science Foundation under Project No. CHE-1213263 and by the Austrian Science Fund (SFB F41, ViCoM). Support was also provided by the Robert A. Welch Foundation under Grant No. D-0005. We are grateful for computer time at the Vienna Scientific Cluster (VSC), project 70376 and to the Fundação de Amparo à Pesquisa do Estado de São Paulo (FAPESP) under Process No. 2013/02972-0 to provide a fellowship to Francisco B. C. Machado for his stay at the Texas Tech University and to Conselho Nacional de Desenvolvimento Científico and Tecnológico (CNPq) for the research fellowship under Process No. 304914/2013-4.

**Table of Contents Graphics**

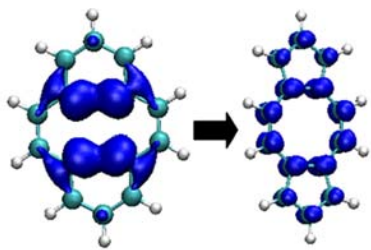

The unpaired density changes from polyradical to closed shell character on geometry relaxation.